\documentclass[reprint,
 amsmath,amssymb,
 aps,
pra,
showkeys
]{revtex4-2}
\usepackage{booktabs}
\usepackage{graphicx}
\usepackage{dcolumn}
\usepackage{bm}
\usepackage{float}
\usepackage{multirow}
\usepackage{natbib}
\usepackage{xcolor}
\usepackage{hyperref}

\begin{document}
\newcommand{\cm}{cm$^{-1}$}
\newcommand{\Aa}{$A_\textrm{g}^1$}
\newcommand{\Ab}{$A_\textrm{g}^2$}
\newcommand{\Ac}{$A_\textrm{g}^3$}
\newcommand{\Ea}{E$\parallel${a}}
\newcommand{\Eb}{E$\parallel${b}}
\newcommand{\EL}{$E_\textrm{L}$}
\newcommand{\CSBC}{CrSBr$_{1-x}$Cl$_{x}$~}


\title{Interplay of Cl Substitution and He$^{+}$ Irradiation in \CSBC}

\author{Satyam Sahu$^1$}
\email{satyam.sahu@jh-inst.cas.cz}
\author{Adeel Bukhari$^{1,2}$}
\author{Arijit Kayal$^1$}
\author{Valerie Černá$^{1,3}$}
\author{Bing Wu$^4$}
\author{Aljoscha Söll$^4$}
\author{Gregor Hlawacek$^5$}
\author{Zdeněk Sofer$^4$}
\author{Martin Kalbáč$^1$}
\author{Matěj Velický$^1$}%
\author{Otakar Frank$^1$}%
\email{otakar.frank@jh-inst.cas.cz}
\affiliation{%
 $^1$J. Heyrovský Institute of Physical Chemistry, Czech Academy of Sciences, Dolejškova 2155/3, 182 23, Prague 8, Czech Republic\\$^2$Faculty of Mathematics and Physics, Charles University, Ke Karlovu 3, 121 16, Prague 2, Czech Republic\\$^3$Department of Physical Chemistry, University of Chemistry and Technology Prague, Technická 5, 166 28 Prague 6, Czech Republic\\$^4$Department of Inorganic Chemistry, University of Chemistry and Technology Prague, Technická 5, 166 28 Prague 6, Czech Republic\\$^5$Institute for Ion Beam Physics and Materials Research, Helmholtz-Zentrum Dresden-Rossendorf, D-01328 Dresden, Germany}%

\date{\today}

\begin{abstract}
Two-dimensional magnetic semiconductors provide a promising platform for exploring the interplay between disorder, lattice dynamics, and resonant light--matter interactions. Among them, CrSBr exhibits strong in-plane anisotropy and pronounced resonance-enhanced Raman scattering. Here, we investigate the effects of Cl substitution and He$^{+}$ irradiation on the vibrational response of CrSBr using polarization-resolved Raman spectroscopy. Cl substitution activates additional phonon modes associated with local symmetry breaking, while He$^{+}$ irradiation introduces distinct defect-related scattering channels and enhanced phonon broadening. The combined effects of alloy disorder and externally introduced defects lead to strong anisotropic reconstruction of the Raman spectra and modification of the nonlinear Raman response under near-resonant 1.96 eV excitation. Power-dependent measurements reveal robust superlinear scaling of both intrinsic and substitution-induced phonon modes, indicating persistent resonance-enhanced electron--phonon coupling even in defect-engineered samples.

\end{abstract}

\maketitle


\section{\label{sec:Introduction}INTRODUCTION}
Low-dimensional van der Waals materials provide an exceptional platform for tuning lattice, electronic, and magnetic properties through composition, dimensionality, and external perturbations \cite{Blundo2021, Chegel2020, Chen2019, Telford2023, Vincent2021}. Among them, CrSBr has recently attracted significant attention as an air-stable layered magnetic semiconductor that combines strong in-plane anisotropy, robust excitonic features, and thickness-dependent interlayer interactions \cite{Ziebel2024, Torres2023, Klein2023, Sahu2025, Lin2024, Antoniazzi2026}. These characteristics make CrSBr particularly appealing for exploring the interplay between structure and functionality in reduced dimensions, as well as for potential applications in optoelectronic and spintronic devices.\par
Beyond pristine compounds, chemical substitution offers an effective route to engineer material properties without disrupting the layered framework \cite{Yang2021, Karthikeyan2019, Sahu2026}. In the CrSBr family, partial replacement of Br by Cl provides access to the mixed-halide series \CSBC, where changes in ionic size, electronegativity, and local bonding environment can modify phonon dynamics, interlayer coupling, electronic structure, and magnetic interactions \cite{Sahu2026, Badola2026, Telford2023}. Such alloying introduces a controllable degree of static disorder while preserving the parent crystal motif, thereby enabling systematic investigation of composition-dependent lattice dynamics \cite{Sahu2026}.\par
A complementary strategy for tailoring layered materials is defect engineering through ion irradiation. Controlled irradiation can generate vacancies, interstitials, antisite defects, and local lattice distortions, offering a post-growth route to modify material properties with spatial and fluence selectivity \cite{Krasheninnikov2026, Ma2025, Long2023, Torres2023, Klein2026}. In 2D materials, irradiation-induced defects may strongly influence phonon lifetimes, strain fields, carrier scattering, and interlayer coupling \cite{Torres2023, Long2023}. Importantly, the response to irradiation is not expected to be universal, but rather governed by the intrinsic bonding landscape, pre-existing disorder, and sample thickness \cite{Long2023}.\par
Raman spectroscopy is particularly well suited for investigating these effects because phonon frequencies, linewidths, intensities, and symmetry-dependent selection rules are highly sensitive to local structure and disorder \cite{Jorio2011}. In layered systems, these signatures can further evolve with thickness and crystal orientation, providing direct insight into dimensionality-dependent defect responses \cite{Torres2023, Sahu2026}. For mixed-halide \CSBC crystals, the combined influence of alloy disorder and irradiation-induced defects remains unexplored.\par
Here, we investigate the composition-dependent irradiation response of \CSBC using thickness-dependent polarized vibrational measurements over a series of controlled irradiation fluences. By comparing multiple alloy compositions across flakes of varying thickness, we reveal how halide substitution modifies the evolution of phonon energies, linewidths, and anisotropic scattering under defect generation. Our results demonstrate that alloying provides an effective handle to tune the defect sensitivity of this layered magnetic semiconductor, establishing mixed-halide compounds as a versatile platform for disorder engineering in low-dimensional materials.
\section{\label{sec:Methods}METHODS}
Single crystals of \CSBC ($0 \leq x \leq 0.5$) were synthesized using established bulk crystal growth procedures, as reported in our previous work \cite{Sahu2026}. Layered crystals with rectangular morphology were mechanically exfoliated prior to optical characterization. Bulk crystals were also retained for measurements requiring a larger scattering volume and improved signal stability.\par
Thin flakes were prepared by mechanical exfoliation of bulk crystals onto SiO$_2$/Si substrates. Candidate flakes were first identified by optical microscopy based on contrast and lateral size. Thicknesses ranging from 3L to 20L regimes were selected across the halide composition series. Because exfoliation yields nonuniform thickness distributions across compositions, each halide alloy was analyzed using a representative set of accessible flakes rather than enforcing identical layer numbers for all samples. The thickness of individual flakes was determined using a combination of optical contrast and atomic force microscopy (AFM).\par
Defects in \CSBC were introduced via He$^{+}$ irradiation using an IonEtch Sputter Gun (Tectra, Germany). The irradiation was performed at an ion energy of 1 keV, with a working pressure of 6 × 10$^{-5}$ mbar and a base pressure below 1 × 10$^{-10}$ mbar. The ion fluence is controlled by varying the irradiation time. To estimate the ion fluence, the ion current was measured under the same pressure conditions. A current of 5 \textmu A was recorded on a 4 cm$^{2}$ collector plate, and the ion density was calculated using $\phi = I/(A e)$, where $I$ is the measured current, $A$ is the collection area, and $e$ is the elementary charge \cite{Park2024}. This yielded an ion fluence of 7.9 × 10$^{12}$ ions cm$^{-2}$ s$^{-1}$. Using this value, the irradiation fluence was determined from the exposure time and ranged from 3.9 × 10$^{13}$ to 1.2 × 10$^{15}$ cm$^{-2}$. Alternatively, a Carl Zeiss Orion NanoFAB He-ion microscope \cite{Hlawacek2014} was used to irradiate selected bulk \CSBC samples with He$^{+}$ ions of 7.5 keV energy. The results presented in the main text figures are solely from samples irradiated using a 1 keV broad beam, while results for He$^+$ microscope irradiated samples are included in the Supplementary Information.\par
Raman measurements were performed using a confocal micro-Raman setup (WITEC Alpha 300R) equipped with a grating of 1800 lines/mm in a backscattering geometry. The continuous-wave laser excitations at 2.33 eV (532 nm) and 1.96 eV (633 nm) were used as the excitation source. The laser beam was focused onto the sample using a high numerical aperture (NA = 0.9) objective, yielding a diffraction-limited spot size of about $0.5$~\textmu m. The spectral resolution was better than $\sim$ 1 \cm , enabling reliable tracking of small peak shifts and linewidth variations. To minimize local heating and photoinduced effects, the incident laser power was kept to 100 \textmu W, except for power-dependent measurements.\\
AFM measurements were performed using a Bruker ICON system operated in PeakForce tapping mode using ScanAsyst Air probes ($k$ = 0.2 - 0.8 N/m). The flake thickness was extracted from a height profile acquired across the flake edges.\\ 
All the measurements were conducted at room temperature (295 K).\\
Raman spectral analysis was performed using Origin 2019. Each Raman spectrum was first normalized with respect to the Raman peak intensity of the Si substrate, followed by subtraction of the substrate spectrum from the corresponding sample spectrum to isolate the Raman response of a material. The resulting spectra were then fitted using a Voigt line-shape function to extract the relevant parameters.
\section{\label{sec:Results}RESULTS}
\begin{figure}[b!]
    \centering
    \includegraphics[width=8.5cm]{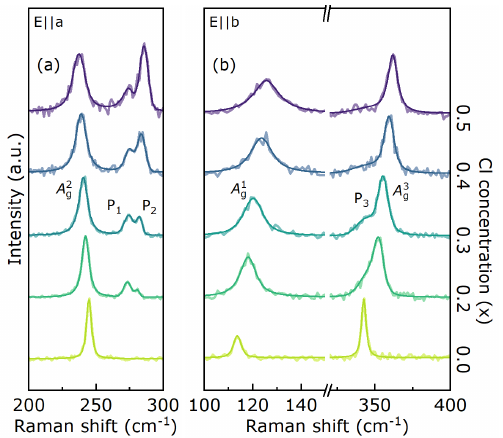}
    \caption{Composition-dependent Raman spectra of bulk \CSBC under 2.33 eV excitation for (a) E$\parallel$a and (b) E$\parallel$b, respectively. Here, the light curves are the raw data, while the darker ones are the fits using a Voigt function.}
    \label{fig:Cl-dependence}
\end{figure}
\begin{figure*}[htbp!]
    \centering
    \includegraphics[width=17.5cm]{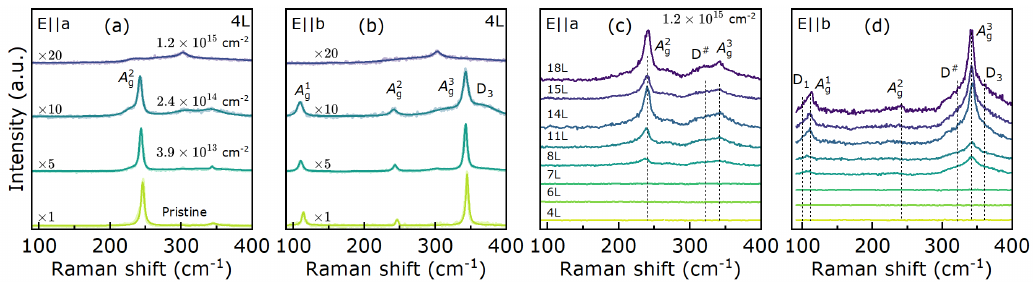}
    \caption{Defects in He$^{+}$ irradiated CrSBr. (a-b) Fluence-dependent Raman spectra of 4L CrSBr, and (c-d) thickness-dependent normalized (w.r.t. Si) Raman spectra at a fluence of $1.2 \times 10^{15}$ cm$^{-2}$ excited using a 2.33 eV laser for \Ea~(a, c) and \Eb~configurations (b, d). The positions of the parent $A_\textrm{g}$ modes and He$^{+}$ induced D$_1$, D$_3$, and D$^{\#}$ modes are highlighted by dotted vertical lines.}
    \label{fig:CrSBr}
\end{figure*}
We first establish the effect of permanent chemical disorder introduced through halide substitution in the \CSBC series. Partial replacement of Br by Cl preserves the layered crystal structure while continuously modifying the local bonding environment, thereby providing a controlled route to tune the intrinsic lattice dynamics through substitutional disorder \cite{Sahu2026, Telford2023}.\par
Figure \ref{fig:Cl-dependence} presents the composition-dependent Raman spectra of bulk \CSBC crystals measured under 2.33 eV excitation for laser and analyser polarized parallel to a- (\Ea; long) and b- (\Eb; short) crystalline axes of CrSBr. Several first-order phonon modes of pristine CrSBr evolve systematically with increasing Cl concentration, confirming the strong sensitivity of lattice vibrations to halogen substitution. The \Aa~mode is primarily associated with halogen vibrations, whereas the \Ab~and \Ac~modes mainly originate from coupled Cr--S vibrations. For \Ea, the prominent \Ab~ mode near 240 \cm~ progressively shifts and changes linewidth with increasing Cl concentration, while additional features labeled P$_1$ and P$_2$ emerge and gain spectral weight for intermediate and high Cl contents. For the case of \Eb, the \Aa~ and \Ac~ modes also show clear composition-dependent renormalization, accompanied by the appearance of the P$_3$ feature. These P-modes arise from local vibrational modifications induced by the substitution of heavier Br atoms with lighter Cl atoms \cite{Sahu2026, Badola2026, Telford2023}.\par
The systematic hardening/softening and broadening of these modes reflect two simultaneous effects of alloying: (i) modification of average force constants and lattice parameters due to Br-to-Cl substitution, and (ii) enhanced phonon scattering caused by static compositional disorder \cite{Sahu2026, Badola2026,Telford2023}. The emergence of the P-modes further indicates local symmetry perturbation and activation of otherwise weak or symmetry-relaxed vibrational channels in the mixed-halide environment \cite{Sahu2026, Telford2023}.\par
A comprehensive discussion of the phonon evolution in \CSBC has been reported in our recent work \cite{Sahu2026}, where mixed-halide alloying was shown to enable continuous tuning of the vibrational properties of this material family. Here, these substituted crystals serve as an ideal platform with a predefined level of permanent disorder. This baseline is essential for the present study, since subsequent irradiation introduces additional dynamic defects on top of a composition-dependent lattice landscape, allowing direct comparison between substitutional and irradiation-induced disorder.\par

Next, we examine the influence of externally introduced defects in pure CrSBr. Figure \ref{fig:CrSBr} presents the Raman response of 4L CrSBr under increasing He$^{+}$ irradiation fluence, together with the thickness evolution measured at the highest fluence of $1.2 \times 10^{15}$ cm$^{-2}$ (also see supplementary figure S1). The measurements reveal a substantial reconstruction of the vibrational landscape following irradiation, manifested through phonon renormalization, anisotropic spectral redistribution, and the activation of additional Raman scattering channels.\par

In the \Ea~configuration (Figure \ref{fig:CrSBr}a), the Raman response remains dominated by the intrinsic phonons of CrSBr. The \Ab~mode persists as the most prominent feature for irradiation fluences up to $2.4 \times 10^{14}$ cm$^{-2}$, with no distinct defect-activated peak appearing within the measured spectral range. Nevertheless, a slight broadening and modulation of the phonon intensity can be observed at intermediate fluences, indicating the onset of irradiation-induced lattice disorder.\par

A markedly different behavior is observed in the \Eb~spectra shown in Figure \ref{fig:CrSBr}b, where an additional broad feature labeled D$_3$ emerges under the \Ac~mode from 300 \cm~to 400 \cm. The intensity of D$_3$ increases with irradiation fluence, confirming its direct association with defect formation \cite{Torres2023, Weile2025}. Previous study on defective CrSBr suggests that this feature originates from intralayer chromium- and sulfur-related defects \cite{Torres2023}. Its pronounced appearance only in the \Eb~configuration highlights the anisotropic nature of the defect-modified vibrational response in CrSBr \cite{Torres2023}.\par

At the highest irradiation fluence, the Raman features in both polarization configurations are strongly suppressed and eventually replaced by a nearly featureless background, indicating severe structural damage in the irradiated thin flake \cite{Hazem2019}.\par

Figures \ref{fig:CrSBr}c-d show the thickness-dependent Raman spectra acquired from CrSBr samples irradiated at a fixed fluence of $1.2 \times 10^{15}$ cm$^{-2}$, with the spectra vertically offset for clarity. In the \Ea~configuration (Figure \ref{fig:CrSBr}c), an additional feature labeled D$^{\#}$ emerges as a lower-frequency shoulder of the \Ac~mode. The \Eb~spectra in Figure \ref{fig:CrSBr}d exhibit a richer defect-related Raman response with thickness, where three additional features labeled D$_1$, D$^{\#}$, and D$_3$ are simultaneously observed. Based on previous studies of irradiated CrSBr, the D$_1$ feature is likely associated with surface-related defects arising from the enhanced sensitivity of the near-surface lattice region to ion-induced disorder, while D$_3$ has been linked to defect-induced chromium and sulphur vacancies in the lattice. \cite{Torres2023, Weile2025}.\par
\begin{figure}[htbp!]
    \centering
    \includegraphics[width=6.5cm]{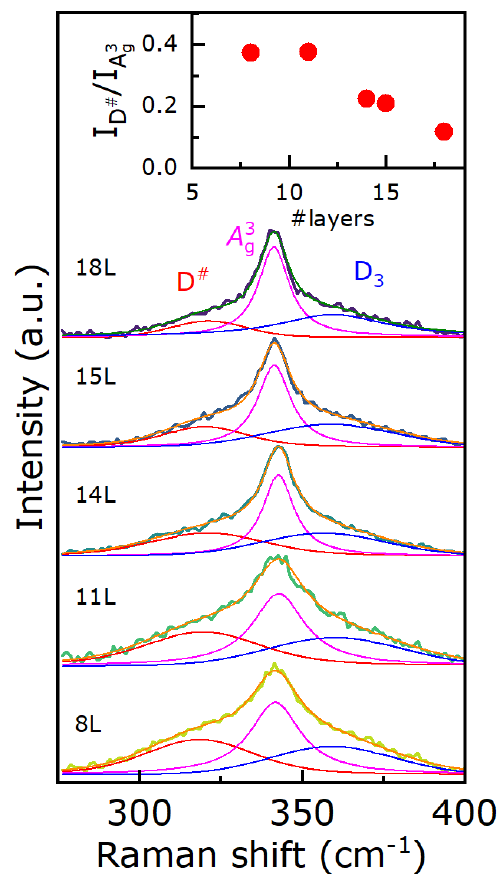}
    \caption{Thickness evolution of D$^{\#}$ and D$_{3}$ modes in CrSBr at a fluence of $1.2 \times 10^{15}$ cm$^{-2}$ under 2.33 eV excitation for E$\parallel$b configuration. Here, the light curves are the raw data, while the darker ones are the fits using the Voigt function for the peaks highlighted using vertical lines in Figure \ref{fig:CrSBr}. Inset: $I_{D^\#}/I_{A_\textrm{g}^3}$ as a function of number of layers.}
    \label{fig:thickness}
\end{figure}

To further understand the origin of the disorder-related D$^{\#}$ feature, we examine its thickness evolution at the highest irradiation fluence ($1.2 \times 10^{15}$ cm$^{-2}$), as shown in Figure \ref{fig:thickness}. The spectra reveal a systematic reduction in the relative spectral weight of the D$^{\#}$ mode with increasing thickness. This trend is quantified in the inset, where the intensity ratio $I_{D^{\#}}/I_{A_\textrm{g}^3}$ decreases monotonically as the flake thickness increases from 8L to 18L. The pronounced enhancement of the D$^{\#}$ contribution in thinner flakes strongly suggests that this mode is associated with surface- or near-surface-related disorder induced during irradiation.\par

In addition to the evolution of the D$^{\#}$ mode, the spectra also exhibit a gradual suppression of the broad D$_3$ contribution with increasing thickness. Since the He$^{+}$ irradiation was performed using relatively low-energy ions (1 keV), the penetration depth of the ions is expected to remain limited to the near-surface region of the crystal \cite{Long2023, Torres2023}. As a result, thinner flakes experience a significantly larger effective defect density throughout their volume, leading to stronger disorder-assisted Raman scattering. In thicker flakes, however, a progressively larger fraction of the crystal remains comparatively undamaged beneath the irradiated surface layer, thereby reducing the relative contribution of the D$_3$ mode. In the bulk sample, the Raman features remain largely unchanged across most irradiation fluences, indicating that the induced damage is predominantly confined to the near-surface region. Only at the highest fluence does the defect-induced D$^{\#}$ mode become discernible, as shown in supplementary figure S6. However, compared to the D$^{\#}$ mode, the thickness-induced suppression of the D$_3$ mode is significantly less pronounced, confirming it originates from the interior of the lattice rather than the surface \cite{Torres2023}.

\begin{figure*}[htbp]
    \centering
    \includegraphics[width=17.5cm]{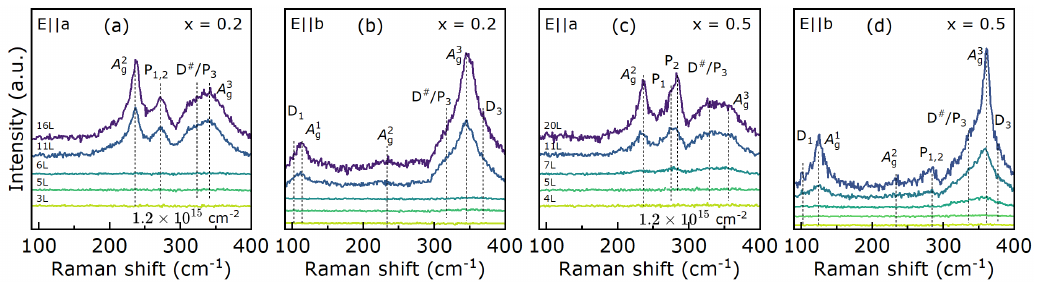}
    \caption{Defects in He$^{+}$ irradiated \CSBC. Thickness-dependent normalized Raman spectra at a fluence of $1.2 \times 10^{15}$ cm$^{-2}$ excited using a 2.33 eV laser for \Ea~and \Eb~configurations for (a-b) $x$ = 0.2 and (c-d) $x$ = 0.5. The positions of the parent $A_\textrm{g}$ modes, Cl-substitution induced P$_1$, P$_2$ and P$_3$ modes, and He$^{+}$ induced D$_1$, D$_3$, and D$^{\#}$ modes are highlighted by dotted vertical lines.}
    \label{fig:CrSBrCl}
\end{figure*}
\begin{figure*}[htbp!]
    \centering
    \includegraphics[width=17.5cm]{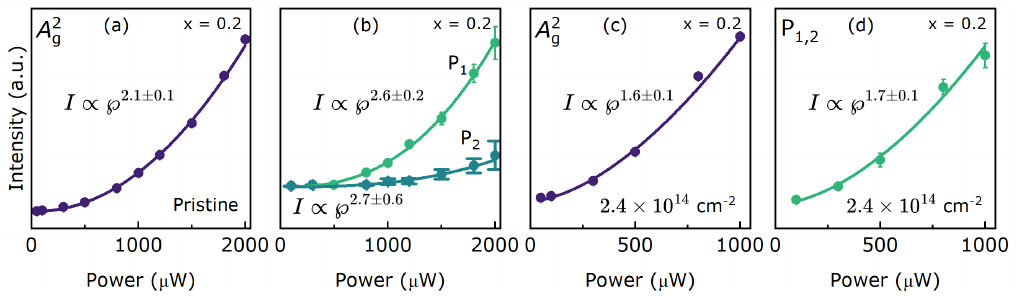}
    \caption{Raman intensities as a function of illumination power under resonant (1.96 eV) excitation for \Ea~configuration for $x$ = 0.2. (a-b)  \Ab, P$_1$, and P$_2$ modes in non-irradiated samples. (c-d) He$^{+}$ irradiated ($2.4 \times 10^{14}$ cm$^{-2}$) \Ab~and P$_{1,2}$-modes. Solid circles are the fitted intensities with the bars as the fitting error, while the curves are the fits using $I \propto \wp^{\theta}$.}
    \label{fig:SRS}
\end{figure*}

The combined influence of substitutional alloying and externally introduced defects on the lattice dynamics of \CSBC is summarized in Figure \ref{fig:CrSBrCl} and supplementary figures S2-S5, which present the polarization-resolved Raman spectra of He$^{+}$-irradiated \CSBC flakes for two different Cl concentrations, $x = 0.2$ and $x = 0.5$, measured as a function of thickness. In contrast to pristine irradiated CrSBr, where the defect response was primarily limited to the emergence of D$_1$, D$_3$, and D$^{\#}$ features, the alloyed samples exhibit a significantly richer spectral reconstruction characterized by the coexistence of substitution-induced and irradiation-activated Raman modes, substantial linewidth broadening, and pronounced redistribution of spectral weight. These observations demonstrate that Cl substitution and He$^{+}$ irradiation cooperatively modify the local lattice symmetry and vibrational landscape of CrSBr.\par
For the lower Cl concentration ($x = 0.2$), the \Ea~spectra shown in Figure \ref{fig:CrSBrCl}a remain dominated by the intrinsic \Ab~phonon. However, compared to irradiated pristine CrSBr, several additional Raman features become visible across the intermediate- and high-frequency regions. In addition to the irradiation-induced D$^{\#}$ and D$_3$ features, the substitution-related P$_1$ and P$_2$ modes, discussed earlier for Cl-alloyed CrSBr \cite{Sahu2026, Telford2023}, are also observed. Unlike the well-resolved splitting reported for non-irradiated alloyed samples, the P$_1$ and P$_2$ modes here appear relatively broad and only partially separated. This reduced spectral resolution likely originates from defect-induced phonon broadening introduced by He$^{+}$ irradiation, which increases phonon scattering and partially overlaps the individual substitution-related modes. A similar broadening behavior is also observed for the nearby intrinsic \Ac~and P$_3$ phonon features. The coexistence of broadened P and D modes, therefore, highlights the strong interplay between alloy disorder and externally introduced defects in reconstructing the Raman response of the material.\par
A substantially stronger spectral reconstruction is observed in the \Eb~configuration for $x = 0.2$ (Figure \ref{fig:CrSBrCl}b). Similar to irradiated pristine CrSBr, the D$^{\#}$ and D$_3$ modes remain strongly enhanced in the \Eb~configuration, reflecting the highly anisotropic nature of defect-assisted Raman scattering in CrSBr. Compared to the pristine compound, the Raman features also exhibit noticeably broader linewidths together with an enhanced spectral background, akin to the increased lattice disorder arising from the combined effects of substitutional alloying and irradiation-induced defects. The robustness and enhancement of the D$_3$ feature indicate that intralayer chromium- and sulfur-related defect complexes \cite{Torres2023} remain stable even in the alloyed system. Moreover, the appearance of both D$^{\#}$ and P$_{3}$ in a similar shift range further hints at the surface-related origin of the D$^{\#}$ Raman mode. A recent work has shown a higher concentration of Br vacancies at the CrSBr sample surfaces \cite{Weile2025}. As P$_{3}$ is closely related to the Cl substitution at Br sites \cite{Sahu2026}, it can be inferred that the P$_{3}$ and D$^{\#}$ share a similar origin, linked to surface halogen defects. \par
The effects become more pronounced upon increasing the Cl concentration to $x = 0.5$, as shown in Figures \ref{fig:CrSBrCl}c-d. In the \Ea~configuration, the substitution-induced P$_1$ and P$_2$ modes gain higher intensity relative to the $x = 0.2$ sample, while the spectral region around the \Ac~phonon evolves into a broadened multi-component structure consisting of D$^{\#}$, P$_3$, and \Ac. The poor spectral resolution of these features reflects significant disorder-induced phonon broadening. Compared to the lower substituted sample, the relative changes introduced by He$^{+}$ irradiation appear less pronounced, particularly for thicker flakes where the Raman response remains comparatively stable. This behavior suggests that the vibrational response at high Cl concentration is already dominated by substitutional disorder, thereby reducing the incremental impact of externally introduced defects. In this regime, irradiation primarily enhances phonon damping and spectral broadening rather than activating distinct additional scattering channels.\par
The \Eb~spectra exhibit a similarly broadened vibrational response with simultaneous observation of D$_1$, P$_3$, and D$_3$. Interestingly, with increased disorder, the D$_3$ feature is much weaker than that of the D$_3$ mode at $x = 0.2$. The absence of substantial spectral reconstruction relative to the $x = 0.2$ sample further supports the interpretation that phonon scattering in \CSBC at high substitution levels is already governed by strong alloy disorder. Consequently, the additional perturbation introduced by He$^{+}$ irradiation leads predominantly to linewidth broadening and intensity redistribution within an already strongly disordered vibrational background.\par

Having established the role of both Cl substitution and He$^{+}$ irradiation on the vibrational landscape of CrSBr and \CSBC, we next investigate how these modifications influence the nonlinear resonant Raman response under near-resonant (1.96 eV) excitation \cite{Sahu2025, Mondal2025}. Previous studies on pristine CrSBr and \CSBC have demonstrated unusually strong resonance-enhanced Raman scattering and stimulated Raman scattering-like behavior near the 1.96 eV excitation energy, originating from strong anisotropic electron--phonon coupling \cite{Sahu2025, Sahu2026}. In the present case, the coexistence of substitutional disorder and irradiation-induced defects provides an opportunity to examine how local symmetry breaking and defect scattering affect the nonlinear phonon amplification processes.\par
Figure \ref{fig:SRS} summarizes the excitation power dependence of Raman mode intensities for the $x = 0.2$ sample before and after He$^{+}$ irradiation. The extracted Raman intensities follow a power-law dependence, $I \propto \wp^{\theta}$, where the exponent $\theta$ characterizes the degree of nonlinear resonant enhancement.\par
For the pristine sample (Figure \ref{fig:SRS}a-b), the intrinsic \Ab~mode exhibits a strongly superlinear dependence with $\theta = 2.1\pm0.1$, significantly exceeding the linear behavior expected for conventional first-order Raman scattering. The substitution-induced P$_1$ and P$_2$ modes display even stronger nonlinear scaling with exponents of $2.6\pm0.2$ and $2.7\pm0.6$, respectively \cite{Sahu2025, Sahu2026}. The enhanced nonlinear response of the P modes indicates that the alloy-induced vibrational channels remain strongly coupled to the resonant electronic transitions. Such behavior further suggests that the local symmetry lowering introduced by Cl substitution creates highly efficient resonance-assisted scattering pathways under near-resonant excitation \cite{Sahu2026}.\par

The evolution of the nonlinear response after He$^{+}$ irradiation is shown in Figures \ref{fig:SRS}c-d, for the sample irradiated at a fluence of $2.4 \times 10^{14}$ cm$^{-2}$. The nonlinear exponent of the \Ab~mode decreases to $1.6\pm0.1$. The reduced exponent suggests that irradiation-induced disorder increases phonon damping and introduces additional scattering pathways that weaken the coherence of the nonlinear Raman process.\par
A similar trend is observed for the substitution-related P mode, where the scaling exponent decreases to $1.7\pm0.1$ after irradiation. In contrast to the non-irradiated sample, the individual P$_1$ and P$_2$ features can no longer be spectrally resolved following irradiation, consistent with the substantial disorder-induced linewidth broadening discussed earlier. Consequently, the Raman response appears as a single broadened P feature. Although the nonlinear enhancement is reduced after irradiation, the persistence of superlinear scaling demonstrates that the resonance-assisted scattering channels remain active even in the presence of moderate defect concentrations. At the highest fluence, however, the spectral features become too broadened to allow reliable fitting. Overall, the results indicate that while He$^{+}$ irradiation suppresses the efficiency of the nonlinear Raman amplification through enhanced disorder and phonon scattering, the underlying resonant electron--phonon coupling in Cl-substituted CrSBr remains remarkably robust.

\section{CONCLUSIONS}\label{sec:Conclusions}
In summary, we investigated the combined effects of Cl substitution and He$^{+}$ irradiation on the vibrational and nonlinear Raman response of CrSBr using polarization-resolved Raman spectroscopy. Cl substitution activates additional alloy-related phonon modes, while He$^{+}$ irradiation introduces defect-induced scattering channels and enhances phonon broadening through increased lattice disorder. The interplay between substitutional disorder and irradiation leads to strong anisotropic reconstruction of the Raman spectra while preserving the characteristic polarization-selective resonant response of CrSBr. Power-dependent measurements further reveal that the nonlinear Raman enhancement remains robust even in the presence of moderate defect concentrations, highlighting the persistence of strong electron--phonon coupling in the alloyed and defect-engineered systems.

\begin{acknowledgments}
\textit{Acknowledgments}-- S.S., A.B., A.K., and O.F. thank the Czech Science Foundation (project No. 23-06174K) for financial support. G.H acknowledges the support of DFG (project No. ER 341/20-1). M.V. acknowledges the support of the Lumina Quaeruntur fellowship No. LQ200402201 by the Czech Academy of Sciences. B.W., A.S., and Z.S. were supported by the ERC CZ program (project LL2101) from the Ministry of Education, Youth and Sports. We thank the Ion Beam Center (IBC) at Helmholtz-Zentrum Dresden-Rossendorf (HZDR) for spatially resolved helium ion irradiation. This work was also supported by the Ministry of Education, Youth, and Sports of the Czech Republic, Project No. CZ.02.01.01/00/22\_008/0004558, co-funded by the European Union, and the CzechNanoLab Research Infrastructure, supported by the Ministry of Education, Youth, and Sports of the Czech Republic (LM2023051).\\
\\
\textit{Data availability}-- The data and analyses underlying this study are available from the HeyRACK repository at [persistent link to data repository TBC].
\end{acknowledgments}




\nocite{*}

\bibliography{bibliography}
\pagebreak
\clearpage
\onecolumngrid
\begin{figure*}[t]
    \centering
    \includegraphics[width=11cm]{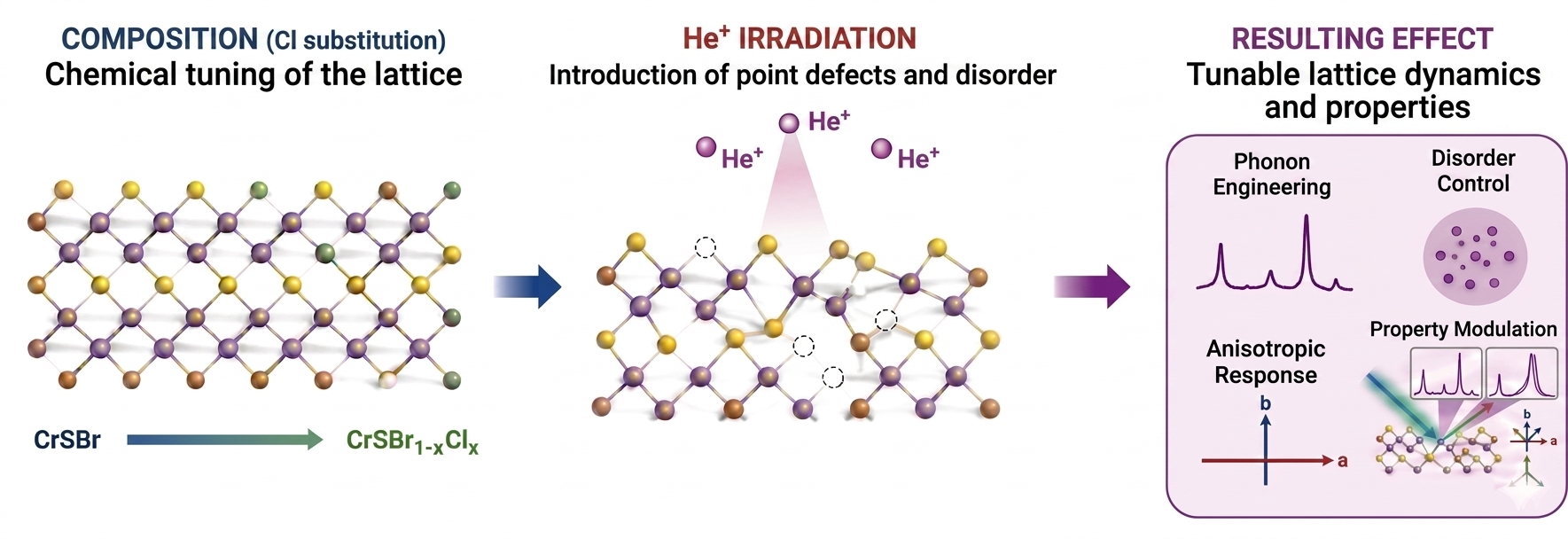}\\
    Table of contents graphic.
\end{figure*}
\pagebreak
\pagebreak
\clearpage
\onecolumngrid
\section*{SUPPLEMENTARY INFORMATION}
\renewcommand{\thefigure}{S\arabic{figure}}
\setcounter{figure}{0}
\renewcommand{\thepage}{S-\arabic{page}}
\setcounter{page}{1}

\begin{figure*}[htbp!]
    \centering
    \includegraphics[width=17.5cm]{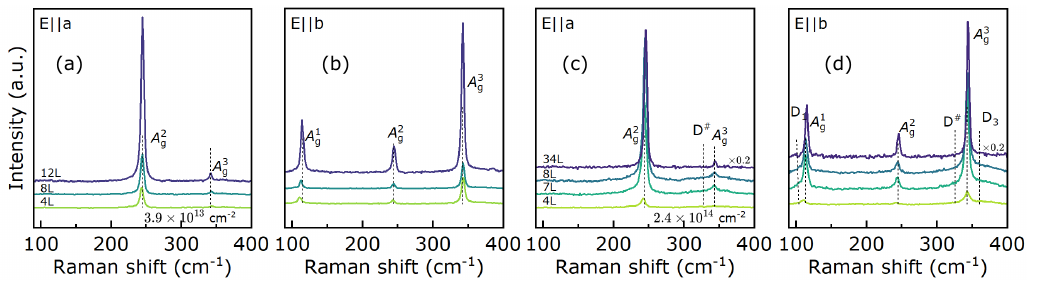}
    \caption{Defects in He$^{+}$ irradiated CrSBr. Thickness-dependent normalized Raman spectra at a fluence of (a-b) $3.9 \times 10^{13}$ cm$^{-2}$ and (c-d) $2.4 \times 10^{14}$ cm$^{-2}$ excited using a 2.33 eV laser for \Ea~and \Eb~configurations. The positions of the parent $A_\textrm{g}$ modes and He$^{+}$ induced D$_1$, D$_3$, and D$^{\#}$ modes are highlighted.}
    \label{fig:CrSBr_SI}
\end{figure*}
\begin{figure*}[htbp!]
    \centering
    \includegraphics[width=17.5cm]{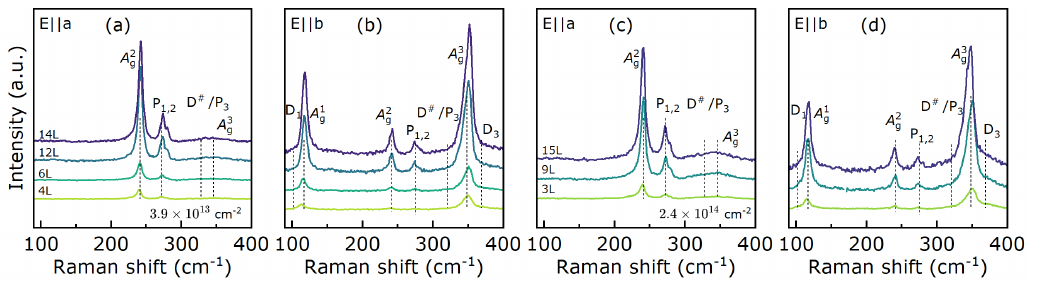}
    \caption{Defects in He$^{+}$ irradiated \CSBC ($x = 0.2$). Thickness-dependent normalized Raman spectra at a fluence of (a-b) $3.9 \times 10^{13}$ cm$^{-2}$ and (c-d) $2.4 \times 10^{14}$ cm$^{-2}$ excited using a 2.33 eV laser for \Ea~and \Eb~configurations. The positions of the parent $A_\textrm{g}$ modes, substitution induced P-modes, and He$^{+}$ induced D$_1$, D$_3$, and D$^{\#}$ modes are highlighted.}
    \label{fig:CrSBr_0.2}
\end{figure*}
\begin{figure*}[htbp!]
    \centering
    \includegraphics[width=17.5cm]{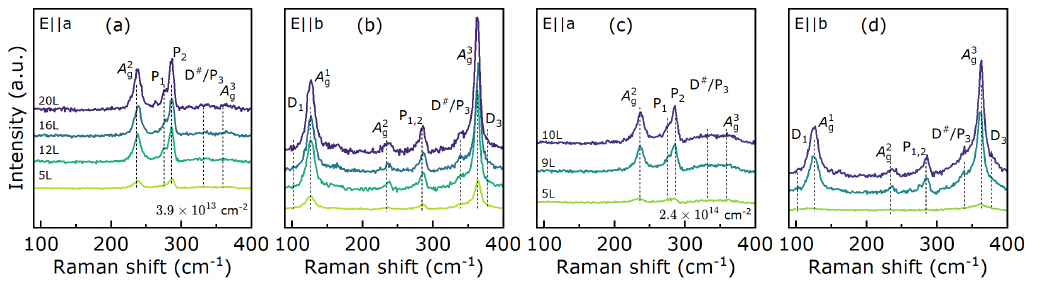}
    \caption{Defects in He$^{+}$ irradiated \CSBC ($x = 0.5$). Thickness-dependent normalized Raman spectra at a fluence of (a-b) $3.9 \times 10^{13}$ cm$^{-2}$ and (c-d) $2.4 \times 10^{14}$ cm$^{-2}$ excited using a 2.33 eV laser for \Ea~and \Eb~configurations. The positions of the parent $A_\textrm{g}$ modes, substitution induced P-modes, and He$^{+}$ induced D$_1$, D$_3$, and D$^{\#}$ modes are highlighted.}
    \label{fig:CrSBr_0.5}
\end{figure*}

\begin{figure*}[htbp!]
    \centering
    \includegraphics[width=8.5cm]{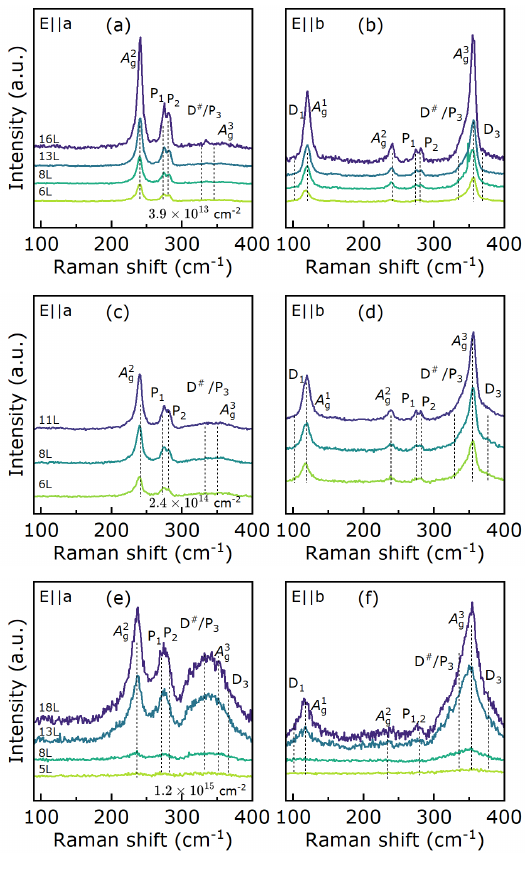}
    \caption{Defects in He$^{+}$ irradiated \CSBC ($x = 0.3$). Thickness-dependent normalized Raman spectra at a fluence of (a-b) $3.9 \times 10^{13}$ cm$^{-2}$, (c-d) $2.4 \times 10^{14}$ cm$^{-2}$ and (e-f) $1.2 \times 10^{15}$ cm$^{-2}$ excited using a 2.33 eV laser for \Ea~and \Eb~configurations. The positions of the parent $A_\textrm{g}$ modes, substitution induced P-modes, and He$^{+}$ induced D$_1$, D$_3$, and D$^{\#}$ modes are highlighted.}
    \label{fig:CrSBr_0.3}
\end{figure*}
\begin{figure*}[htbp!]
    \centering
    \includegraphics[width=8.5cm]{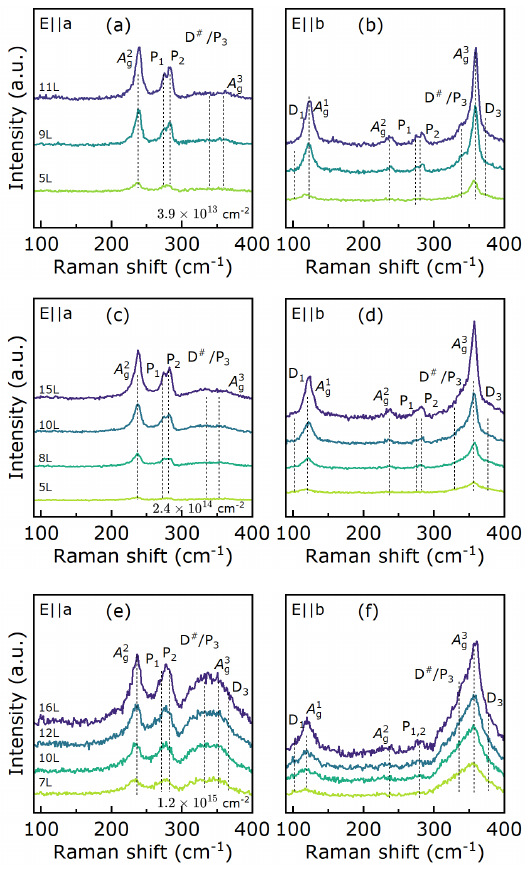}
    \caption{Defects in He$^{+}$ irradiated \CSBC ($x = 0.4$). Thickness-dependent normalized Raman spectra at a fluence of (a-b) $3.9 \times 10^{13}$ cm$^{-2}$, (c-d) $2.4 \times 10^{14}$ cm$^{-2}$ and (e-f) $1.2 \times 10^{15}$ cm$^{-2}$ excited using a 2.33 eV laser for \Ea~and \Eb~configurations. The positions of the parent $A_\textrm{g}$ modes, substitution induced P-modes, and He$^{+}$ induced D$_1$, D$_3$, and D$^{\#}$ modes are highlighted.}
    \label{fig:CrSBr_0.4}
\end{figure*}
\begin{figure*}[htbp!]
    \centering
    \includegraphics[width=17.5cm]{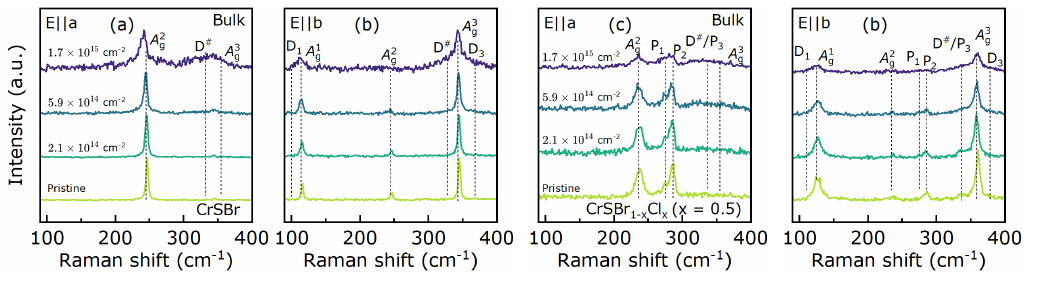}
    \caption{Fluence-dependent Raman spectra of bulk (a-b) CrSBr and (c-d) \CSBC (x = 0.5) acquired using a 2.33 eV excitation after irradiation with designated He$^{+}$ fluences using a helium ion microscope (HIM). The positions of the parent $A_\textrm{g}$ modes, substitution-induced P-modes, and He$^{+}$ induced D$_1$, D$_3$, and D$^{\#}$ modes are highlighted.}
    \label{fig:HIM}
\end{figure*}

\end{document}